\documentclass[bm,aps,showpacs,amsfonts,amssymb]{revtex4}  

\usepackage{graphicx}
\usepackage{natbib}

\usepackage{color}
\font\mybbsub=msbm10 at 8pt
\font\mybbsmall=msbm10 at 10pt

\def\bbsub#1{\hbox{\mybbsub#1}}
\def\bbsmall#1{\hbox{\mybbsmall#1}}

\def\RRsub {\bbsub{R}}
\def\RRsmall {\bbsmall{R}}

\newcommand\beqa{\begin{eqnarray}}
\newcommand\eeqa{\end{eqnarray}}
\newcommand\n{\nonumber\\}
\begin{document}

{~}

\title{``Flip" of $SL(2,R)$-duality in five-dimensional supergravity }
\vspace{2cm}
\author{Shun'ya Mizoguchi\footnote{E-mail:mizoguch@post.kek.jp} and Shinya Tomizawa\footnote{E-mail:tomizawa@post.kek.jp}}
\vspace{2cm}
\affiliation{
Theory Center, Institute of Particle and Nuclear Studies,
KEK, Tsukuba, Ibaraki, 305-0801, Japan 
}
\begin{abstract} 
The dimensional reduction of the bosonic sector of five-dimensional minimal supergravity to a Lorentzian four-dimensional spacetime leads to a theory with a massless axion  
and a dilaton coupled to gravity and two $U(1)$ gauge fields and the dimensionally reduced equations of motion have $SL(2,\RRsmall)/SO(2)$-duality invariance. 
In our previous work, utilizing the duality invariance, we formulated solution-generation techniques within five-dimensional minimal supergravity. 
In this work, 
by choosing a timelike Killing vector, we consider dimensional reduction to a four-dimensional Euclidean space, in which 
the field equations have $SL(2,\RRsmall)/SO(1,1)$ invariance. 
 In the timelike case, we develop a new duality transformation technique, 
 while in the spacelike case we have done that in the previous work. As an example, by applying it to the Rasheed solutions, we obtain rotating Kaluza-Klein black hole solutions in five-dimensional minimal supergravity. In general, in contrast to the spacelike case, the resulting dimensionally reduced solution includes the so-called NUT parameter and therefore from a four-dimensional point of view, such a spacetime is not asymptotically flat. However, it is shown that in some special cases, it can describe ordinary Kaluza-Klein black holes.
\end{abstract}

\preprint{KEK-TH 1466}
\pacs{04.50.+h  04.70.Bw\\
KEK-TH 1515}
\date{\today}
\maketitle

\section{Introduction}\label{sec:intro}

In modern string/supergravity theories and gauge theories,  %and various related contexts, 
higher dimensional black holes and other extended black objects have played important roles. In particular, physics of
black holes in the five-dimensional Einstein-Maxwell-Chern-Simons (EMCS) theory has recently been one of the subjects of 
increasing interest 
%%%%%
since the discovery of the black ring \cite{ER,Pom,MishimaIguchi,EEMR,EEMR2,Y3} and other black object solutions with multiple horizons~\cite{diring,saturn,Izumi,bi,EEF}. 
%%%%%
The five-dimensional EMCS
theory describes the bosonic sector of five-dimensional minimal supergravity, 
which is one of the simplest supergravity sharing many common features with the eleven-dimensional  
supergravity, and can be obtained as a certain low-energy limit of compactified string theory.
%and can also be considered as the simplest and most fundamental one in higher dimensional supergravity theories.
In particular, hidden symmetries such dimensionally reduced theories possess are of technical importance since they enables us to construct non-linear sigma models~\cite{MO,BCCGSW}, which can be useful tools for the proofs of black hole uniqueness theorems~\cite{TYI,TYI2,T} or for solution-generation of black holes. In fact, various types of black-hole solutions in the EMCS theory have so far been found, with the help of the solution-generating techniques recently developed by using such non-linear sigma models~\cite{BCCGSW,GRS,CBJV,CBSV,TYM,GS1,GS2}.

%So far, various types of black hole solutions in  five-dimensional theories have been derived with the help of recent development of solution-generation techniques, such as non-linear sigma model 
%approach~\cite{Rasheed,FGPS,Giusto-Saxena,CEFGS,GRS,BCCGSW,CBJV,TYM,GS1,GS2}, 
%as well as supersymmetric black hole solutions
%~\cite{Gauntlett0,BMPV,Elvang}. 

\medskip

The dimensional reduction of the bosonic sector of five-dimensional minimal supergravity to four dimensions leads to a theory with a massless axion  
and a dilaton coupled to gravity and two $U(1)$ gauge fields with Chern-Simons coupling  \cite{CN,MO,Germar,MizoguchiAsymObfd}. 
As was shown in Ref.~\cite{MO}, the field equations derived by the dimensional reduction are invariant under the action of a global $SL(2,\RRsmall)$ group, by which Maxwell's fields are related to Kaluza-Klein's electromagnetic fields.
This so-called $SL(2,\RRsmall)$-duality invariance enables us to generate a new solution in (the bosonic sector of) five-dimensional minimal supergravity by stating from a certain known solution in the same theory, as the $G_{2(+2)}$-duality invariance \cite{MO} does. 
In our previous work~\cite{MT}, we utilized a spacelike Killing vector for the dimensional reduction (hence the dimensionally reduced four-dimensional space is Lorentzian) and 
developed a formulation in which by using
a certain known solution in five-dimensional pure gravity as a seed solution 
one can obtain new solutions in five-dimensional minimal supergravity.
%The dimensional reduction of the Lagrangian can be done mostly in parallel in both timelike and spacelike cases, except the sign flips of some terms which 
%result in different coset spaces, $SL(2,\RRsmall)/SO(2)$ if spacelike and $SL(2,\RRsmall)/SO(1,1)$ 
%if timelike, just like the $SL(2,\RRsmall)$ symmetries of dimensionally reduced pure Einstein theory \cite{Ehlers,MM,BMG}.   

%
\medskip
One reason for our interest in developing new solution-generating techniques is the 
possibility that they might be used to generate the most general KK black hole solutions 
in five-dimensional minimal  supergravity. In \cite{Tomizawa} it has been shown that such a 
solution is characterized by six parameters - the mass, angular momentum and 
electric/magnetic charges of the Kaluza-Klein gauge field and Maxwell field, respectively. 
So far, ones with some charges of them have been discovered by several authors~\cite{DM,Rasheed,IM,NIMT,TIMN,TI,T2,Gauntlett0,Gaiotto,Elvang3}.
In \cite{MT} we applied the spacelike-Killing $SL(2,\RRsmall)$ transformation to the 
Rasheed solutions, which are known to describe dyonic rotating black holes 
(from the four-dimensional point of view) of five-dimensional pure gravity and 
has four independent parameters, and successfully obtained a new class of 
KK black-hole solutions with {\em five} independent parameters, though we were not be able to find,
within that framework, 
the ones with the maximal number ($=6$) of parameters.

\medskip

Therefore, in this paper, we focus on another $SL(2,\RRsmall)$ symmetry which appears 
in a spacetime with a {\em timelike} Killing vector. 
%%%%%
The dimensional reduction of the Lagrangian can be done mostly in parallel in both timelike and spacelike cases, except the sign flips of some terms which 
result in different coset spaces, $SL(2,\RRsmall)/SO(2)$ if spacelike and $SL(2,\RRsmall)/SO(1,1)$ 
if timelike, just like the $SL(2,\RRsmall)$ symmetries of dimensionally reduced pure Einstein theory \cite{Ehlers,MM,BMG}.   
%%%%%
This $SL(2,\RRsmall)$ is, of course, 
a subgroup of the $G_{2(+2)}$ symmetry if the spacetime allows another 
spacelike Killing vector, and the two  $SL(2,\RRsmall)$ group actions do not commute 
but generate the whole $G_{2(+2)}$. 

%Thus we are motivated to apply
% this {\em timelike}-Killing $SL(2,\RRsmall)$ transformation to five-parameter solutions 
%previously obtained to find new six-parameter solutions.

\medskip

The purpose of this paper is to 
examine whether or not this timelike-Killing $SL(2,\RRsmall)$ can be used for 
generating six-parameter solutions. We will see that it cannot, unfortunately. 
We first
present the $SL(2,\RRsmall)$ transformation formulas in a unified way so that they can be used in both spacelike and timelike cases if the signs are appropriately chosen, and generalize %the solution-generating technique in 
our previous work~\cite{MT} to 
%so that it  can also be used in 
the timelike case. 
Then, 
as an example, we apply it to the Rasheed solutions again to
obtain rotating Kaluza-Klein black hole solutions in five-dimensional minimal supergravity. 
%while in previous work~\cite{MT}, this has been done in the spacelike case. 
As will be shown later, in general the resulting (dimensionally reduced) spacetime geometry (which can be derived by the flip) has the so-called NUT parameter, and hence the four-dimensional reduced spacetime is not an asymptotic Minkowski spacetime. However, in some special cases, it can be shown that the NUT parameter vanishes and hence in that case, it can describe usual Kaluza-Klein black holes, {\it i.e.,} asymptotically flat black holes from a four-dimensional point of view.

\medskip
The remainder of this paper is organized as follows: 
In the next section, %as is done for a spacelike Killing vector in our previous paper, 
we will 
discuss our strategy for the solution-generation technique 
in both the spacelike and timelike cases.
In particular, we will show that in the timelike case, the dimensionally reduced field equations have $SL(2,\RRsmall)/SO(1,1)$ invariance. 
In Section III, by acting the $SL(2,\RRsmall)$ transformation on a certain seed solution, we write down some necessary formulas.
In Section IV, we provide some brief review concerning the Rasheed solution, which we use as a seed in this paper. 
Also, applying actually this formalism to the Rasheed solutions, we present black hole solutions and %furthermore, 
study some basic properties, in particular, its asymptotics.
Section V is devoted to summarizing our results and discussing our new method. In Appendix A, we
apply the duality transformation to the asymptotically flat 
five-dimensional Myers-Perry black holes~\cite{MP}, and show that the solution obtained thereby coincides with the Cveti{\v c}-Youm solution~\cite{CY96}. In Appendix B, we clarify the precise relationship between the $SL(2,\RRsmall)$-duality transformation and the non-linear sigma model approach provided in~\cite{BCCGSW}.

\section{D=4 SL(2,\RRsub) duality with a timelike Killing}
In this section, we summarize the $SL(2,\RRsmall)$ duality symmetry 
of five-dimensional minimal supergravity in the presence of a timelike Killing vector 
field. The reduction procedure is basically the same as
the spacelike case, except the sign flips of some terms in the Lagrangian and 
the duality relations. They 
result in different coset spaces, $SL(2,\RRsmall)/SO(2)$ if spacelike and $SL(2,\RRsmall)/SO(1,1)$ 
if timelike, just like Ehlers' or Matzner-Misner's $SL(2,\RRsmall)$ symmetry of dimensionally 
reduced pure Einstein theory. 
% In Appendix we describe the details of 
%the derivation, where we also discuss the relation to the Harrison transformation 
%in the $G_{2}$ sigma model approach.

\medskip

The conventions and notations are basically the same as those used in \cite{MT}.
The Lagrangian is  
\beqa
{\cal L}
=E^{(5)}\left(R^{(5)}-\frac14 F_{MN}F^{MN}\right)
-\frac1{12\sqrt{3}}\epsilon^{MNPQR}F_{MN}F_{PQ}A_R.
\label{5DL}
\eeqa
The vielbein $E^{(5)A}_{~M}$ is related 
to the five-dimensional metric $G^{(5)}_{MN}$ as
\beqa
G^{(5)}_{MN}&=&E^{(5)A}_{~M}E^{(5)B}_{~N}\eta_{AB},
\eeqa
where 
\beqa
\eta_{AB}=\mbox{diag}(+1,+1,+1,+1,-1)
\eeqa
in the present case. 
%$E^{(5)}=\sqrt{-G^{(5)}}$ is the determinant of $E^{(5)A}_{~M}$. 
$\epsilon^{MNPQR}$ is the densitized anti-symmetric tensor which takes 
values $\pm1$.

\medskip

As usual, we 
decompose the 
vielbein and gauge field as
\beqa
E^{(5)A}_{~M}&=&\left(
\begin{array}{cc}
\rho^{-\frac12} E^{(4)\alpha}_{~~\mu}
&B_\mu \rho \\
0& \rho
\end{array}
\right),\\
A_M&=&(A_\mu, A_t),
\eeqa
where
$\mu$ ($\alpha$) is the four-dimensional curved (flat) index.
Correspondingly, the five-dimensional coordinates $x^M$ are grouped 
into $(x^\mu, t)$, and 
$\eta_{\alpha\beta}=
 \delta_{\alpha\beta}
$. 
All the fields are assumed to be independent of $t$.

\medskip

To make the $D=4$ $SL(2,\RRsmall)$ symmetry manifest, we need to 
dualize the Maxwell field $A_\mu$ \cite{CJ,MO} into
\beqa
\tilde{A}_{\mu\nu}&=&-\rho(\ast F^{(4)})_{\mu\nu}-\frac 2{\sqrt 3}A_t F^{(4)}_{\mu\nu}
+\frac1{\sqrt 3}A_t^2 B_{\mu\nu},
\eeqa
where
\beqa
F^{(4)}_{{\mu}{\nu}}
 &\equiv& F'_{{\mu}{\nu}} + B_{{\mu}{\nu}}A_t,\\
F'_{{\mu}{\nu}}
&\equiv&\partial_{{\mu}}A'_{{\nu}}-\partial_{{\nu}}A'_{{\mu}},\\
A'_{{\mu}}&\equiv&A_{{\mu}}-B_{{\mu}} A_t,\label{eq:A'},\\
B_{{\mu}{\nu}} &\equiv&  \partial_{{{\mu}}} B_{{\nu}}- \partial_{{{\nu}}} B_{{\mu}}.
\eeqa
%Note that the kinetic terms of the scalar $A_y$ and the Kaluza-Klein vector field 
%$B_\mu$ become wrong-sign when $\xi=-1$ (that is, the timelike Killing case).
We also define
\beqa
{\cal G}_{{\mu}{\nu}}
&\equiv&\left(\begin{array}{c}\tilde{A}_{{\mu}{\nu}}\\
B_{{\mu}{\nu}}\end{array} \right)
%\\ 
%{\tilde A}_{{\mu}{\nu}}&\equiv& \partial_{{{\mu}}}{\tilde A}_{{\nu}}-\partial_{{{\nu}}}{\tilde A}_{{\mu}}
\eeqa 
and
\beqa
{\cal H}_{\mu\nu}~\equiv~\left(
\begin{array}{c}
{\cal H}_{\mu\nu}^{\tilde{A}}\\
{\cal H}_{\mu\nu}^B
\end{array}
\right)
&\equiv&m(\ast {\cal G})_{\mu\nu} + a~{\cal G}_{\mu\nu},
\label{H_munu}
\eeqa
where
%$N^{{\mu}{\nu}{\rho}{\sigma}}$ is a two-by-two matrix 
%defined as follows:
\beqa
%N^{{\mu}{\nu}{\rho}{\sigma}}
%&=& m~{1}^{{\mu}{\nu}{\rho}{\sigma}}
%+a~(\ast)^{{\mu}{\nu}{\rho}{\sigma}},\n
V^{-1}mV^{-1}&=&K-\frac12(\Phi\Phi^*K+K\Phi^*\Phi)
+\frac14\Phi\Phi^{*2}K\Phi,\n
V^{-1}aV^{-1}&=&-\Phi^*K+\Phi+\frac12(\Phi\Phi^{*2}K+K\Phi^{*2}\Phi)
 -\frac13\Phi\Phi^*\Phi-\frac14\Phi\Phi^{*3}K\Phi,\\
 V &\equiv& \left(\begin{array}{cc}\rho^{-\frac12}& 0 \\
 0 &\rho^{\frac32}\end{array}
\right),~~
 \Phi\equiv\left(\begin{array}{cc}
 0 &\sqrt{3}\phi\\
 \sqrt{3}\phi& 0 \end{array}
\right),~~
\Phi^*\equiv\left(\begin{array}{cc} 2\phi& 0 \\ 0 &0\end{array}
\right),\\
K &\equiv&(1+\Phi^{*2})^{-1},~~\phi\equiv\frac1{\sqrt{3}}\rho^{-1}A_t.
\eeqa
Explicitly,
\beqa
{\cal H}_{\mu\nu}^{\tilde{A}}&=&A_t B_{\mu\nu}-F^{(4)}_{\mu\nu}~=~-F'_{{\mu}{\nu}},\label{eq:HA}\\
{\cal H}_{\mu\nu}^{B}&=&\frac{A_t^3 B_{\mu\nu}}{3 \sqrt{3}}-\frac{A_t^2 F^{(4)}_{\mu\nu}}{\sqrt{3}}
-\rho A_t \ast F^{(4)}_{\mu\nu}+\rho^3\ast B_{\mu\nu}.\label{eq:HB}
\eeqa 
Note that the sign of second term of (\ref{H_munu}). 
Also, two of the terms of $a$ has changed their signs compared to the spacelike case 
\cite{MT},
while the matrix $m$ has remained unchanged.

\medskip

It is convenient to introduce 
\beqa
{\cal F}_{\mu\nu}&\equiv&\left(
\begin{array}{c}
{\cal G}_{\mu\nu}\\
{\cal H}_{\mu\nu}
\end{array}
\right),
\eeqa
so that the equations of motion and the Bianchi identities are expressed 
in a unified way: 
\beqa
d{\cal F}&=&0,~~~{\cal F}\equiv\frac12 {\cal F}_{\mu\nu} dx^\mu\wedge dx^\nu.
\eeqa
This means that ${\cal F}$ must be written as $d{\cal A}$ for some four-component column
gauge potential vector ${\cal A}={\cal A}_{\mu}dx^\mu$, where
\beqa
{\cal A}_\mu&\equiv&\left(
\begin{array}{c}
{\tilde A}_\mu\\
B_\mu\\
-A'_\mu\\
{\cal H}^B_\mu
\end{array}
\right),
\eeqa
where
${\tilde A}_\mu$ and ${\cal H}^B_\mu$ are some gauge potentials that 
satisfy ${\tilde A}_{\mu\nu}=\partial_\mu{\tilde A}_\nu-\partial_\nu{\tilde A}_\mu$
and 
${\cal H}^B_{\mu\nu}=\partial_\mu{\cal H}^B_\nu-\partial_\nu{\cal H}^B_\mu$,
respectively.

\medskip

It can be shown that ${\cal F}_{\mu\nu}$ satisfies
\beqa
%{\cal F}_{\mu\nu}&=&
%\Omega ~\tau({\cal V}^{-1})  {\cal V}(\ast{\cal F})_{\mu\nu},\\
{\cal F}_{\mu\nu}&=&
{\cal V}^{-1} \Omega  {\cal V}(\ast{\cal F})_{\mu\nu},
\eeqa
where
\beqa
{\cal V}={\cal V}_- {\cal V}_+ ,
~~~~~
{\cal V}_+=\left(
\begin{array}{cc}
~V~~&\\
&V^{-1}
\end{array}
\right),
\quad{\cal V}_-=\exp\left(
\begin{array}{cc}
&-\Phi^*\\
-\Phi&
\end{array}
\right),
~~~
\Omega=
\left(
\begin{array}{cccc}
&&1~&~\\
&&~&1\\
~1~&~~&&\\
~~&~1~&&
\end{array}
\right).
\eeqa
The scalar Lagrangian ${\cal L}_S$ can be written, using 
\beqa
{\cal R}=\Omega{\cal V}^{-1}\Omega{\cal V}, \label{Rmatrix}
\eeqa
or %more simply 
\beqa
{\cal R}'={\cal V}^{-1}\Omega{\cal V}={{\cal R}'}^{-1},\label{R'matrix}
\eeqa
as
\beqa
{\cal L}_S
&=&\frac3{40} E^{(4)} {\rm Tr}\partial_{{\mu}}{{\cal R}'}
\partial^{{\mu}}{\cal R}'.
%\n
%&=&-\frac3{40} E^{(4)} {\rm Tr}\left
%(\partial_{{\mu}}{\cal V}\cdot{\cal V}^{-1}-\tau(\partial_{{\mu}}{\cal V}\cdot{\cal V}^{-1})
%\right)^2\label{L_S}
\eeqa
The equations of motion and the Bianchi identity are invariant under
\beqa
{\cal F}_{\mu\nu}&\mapsto&\Lambda^{-1}{\cal F}_{\mu\nu},\label{eq:Ftr}\\
{\cal V} &\mapsto& {\cal V} \Lambda, \label{VVtoVVU-1}\\
%\eeqa
%and the scalar Lagrangian ${\cal L}_S$ (\ref{L_S}) is also manifestly invariant.
%${\cal R}$ transforms as
%%\beqa
{\cal R}'&\mapsto& \Lambda^{-1}{\cal R}' \Lambda.
\label{R'tr}
\eeqa

Since the $SL(2,\RRsmall)$ transformation (\ref{eq:Ftr}) is a rigid one, 
it can also be written at the gauge potential level:
\beqa
{\cal A}_\mu&\mapsto&\Lambda^{-1}{\cal A}_\mu.
\eeqa
The point is that, once the gauge potential vector ${\cal A}_\mu$ can be 
computed for the seed solution, the new $B_\mu$ and $-{A'}_\mu$ fields 
can be found by simply a trivial matrix multiplication. The nontrivial task is 
the computation of ${\tilde A}_\mu$ and ${\cal H}^B_\mu$, but the effort is 
reduced compared with the integrations of the potentials 
in the $G_2$ sigma model approach.

\medskip

The scalar matrix ${\cal R}$ is defined as a $4\times 4$ matrix in (\ref{Rmatrix}); it is more 
convenient to consider scalars in the defining representation of $SL(2,\RRsmall)$ 
by using the Lie algebra isomorphism $\pi$:
\beqa
\pi(E')=\left(
\begin{array}{cc}0&1\\0&0
\end{array}
\right),~~
\pi(F')=\left(
\begin{array}{cc}0&0\\1&0
\end{array}
\right),~~
\pi(H')=\left(
\begin{array}{cc}1&0\\0&-1
\end{array}
\right).\label{isomorphism}
\eeqa
A generic group element of $SL(2,\RRsmall)$ can be expressed as 
\beqa
\Lambda=e^{-\delta E'}e^{(\log\gamma) H'}e^{-\epsilon F'}
\eeqa
\label{Lambda}
which corresponds to 
\beqa
\pi(\Lambda)&=&
\left(
\begin{array}{cc}
1&-\delta\\0&1
\end{array}
\right)
\left(
\begin{array}{cc}
\gamma&0\\0&\gamma^{-1}
\end{array}
\right)
\left(
\begin{array}{cc}
1&0\\ -\epsilon&1
\end{array}
\right)
~=~
\left(
\begin{array}{cc}
a&b\\c&d
\end{array}
\right)
\eeqa
for nonzero $d$, 
where $a$, $b$, $c$, $d$, $\delta$, $\epsilon$ and $\gamma$ are all real numbers 
with $ad-bc=0$ and $\gamma\neq 0$. 
The element with $d=0$ can be obtained by blowing up 
the singularity that occurs in the limit $\delta\rightarrow 0$. 
Since the Cartan algebra degree of freedom does not add a new parameter 
to the solutions, we set $\gamma=1$ for simplicity and ignore the $e^{H'}$ factor \cite{MT}. 
(In fact, it also turns out 
that this $SL(2,\RRsmall)$ transformation can add only one independent parameter, 
and $\delta$ may also be set to zero \cite{MT}.) Hence
\beqa
\Lambda&=&\left(
\begin{array}{cccc}
 3 \delta ^2 \epsilon ^2+4 \delta  \epsilon +1 & \sqrt{3} \delta ^2 &
   -\delta  (3 \delta  \epsilon +2) & -\sqrt{3} \epsilon  (\delta 
   \epsilon +1)^2 \\
 \sqrt{3} \epsilon ^2 & 1 & -\sqrt{3} \epsilon  & -\epsilon ^3 \\
 -\epsilon  (3 \delta  \epsilon +2) & -\sqrt{3} \delta  & 3 \delta 
   \epsilon +1 & \sqrt{3} \epsilon ^2 (\delta  \epsilon +1) \\
 -\sqrt{3} \delta  (\delta  \epsilon +1)^2 & -\delta ^3 & \sqrt{3} \delta
   ^2 (\delta  \epsilon +1) & (\delta  \epsilon +1)^3
\end{array}
\right),
\label{Lambda_gamma=0}\\
\Lambda^{-1}&=&
\left(
\begin{array}{cccc}
 3 \delta  \epsilon +1 & \sqrt{3} \delta ^2 (\delta  \epsilon +1) &
   \delta  (3 \delta  \epsilon +2) & \sqrt{3} \epsilon  \\
 \sqrt{3} \epsilon ^2 (\delta  \epsilon +1) & (\delta  \epsilon +1)^3 &
   \sqrt{3} \epsilon  (\delta  \epsilon +1)^2 & \epsilon ^3 \\
 \epsilon  (3 \delta  \epsilon +2) & \sqrt{3} \delta  (\delta  \epsilon
   +1)^2 & 3 \delta ^2 \epsilon ^2+4 \delta  \epsilon +1 & \sqrt{3}
   \epsilon ^2 \\
 \sqrt{3} \delta  & \delta ^3 & \sqrt{3} \delta ^2 & 1
\end{array}
\right),
\eeqa
\beqa
\pi(\Lambda)
&=&
\left(
\begin{array}{cc}
 \delta  \epsilon +1 & -\delta  \\
 -\epsilon  & 1
\end{array}
\right).
\eeqa
Also we find
\begin{eqnarray}
\pi({\cal R})&=&
\left(
\begin{array}{cc}
\rho^{-1}& -\frac{1}{\sqrt{3}}\rho^{-1}A_t  \\
+\frac{1}{\sqrt{3}}\rho^{-1}A_t & - \frac{1}{3}\rho^{-1}A_t^2+\rho
\end{array}
\right),
%\\
%\to \pi({\cal R}^{new})=\pi(\Lambda^T {\cal R}\Lambda) 
%&=&\left(
%\begin{array}{cc}
%\left(1+\beta(\alpha+\frac{ A_5}{\sqrt{3}})\right)^2\rho^{-1}+\beta^2\rho & -\left(1+\beta(\alpha+\frac{A_5}{\sqrt{3}})\right)\left(\alpha+\frac{A_5}{\sqrt{3}}\right)\rho^{-1}-\beta\rho\\
%-\left(1+\beta(\alpha+\frac{A_5}{\sqrt{3}})\right)\left(\alpha+\frac{A_5}{\sqrt{3}}\right)\rho^{-1}-\beta\rho & (\alpha+\frac{A_5}{\sqrt{3}})^2 \rho^{-1}+\rho\label{eq:trR}
%\end{array}
%\right).\n
\end{eqnarray}
which transforms as
\beqa
\pi({\cal R})&\mapsto&
%\pi(\tau(\Lambda))^{-1}\pi({\cal R})\pi(\Lambda).\\
\pi(\Omega\Lambda^{-1}\Omega)\pi({\cal R})\pi(\Lambda),~~~~~
\pi(\Omega\Lambda^{-1}\Omega)=\left(
\begin{array}{cc}
 \delta  \epsilon +1 & \epsilon   \\
 \delta & 1
\end{array}
\right).
\label{2Dscalartransformation}
\eeqa

\medskip

\section{Transformation Formulas}
The $SL(2,\RRsmall)$-duality transformation requires that solutions should admit the 
existence of at least a single Killing isometry.  In this paper, we assume that a spacetime admits two commuting Killing vector fields, timelike one $\partial/\partial t$ (at least at infinity) and spacelike one $\partial/\partial x^5$ and that each component of the spacetime metric and gauge potential $1$-form are independent of $t$ and $x^5$.  
While in our previous work~\cite{MT}  
we have used the spacelike Killing vector $\partial/\partial x^5$
for $SL(2,\RRsmall)$-duality transformation, where the five-dimensional metric is written as 
\begin{eqnarray}
ds^2=\rho^2(dx^5+B_\mu dx^\mu)^2+\rho^{-1}ds^2_{(4)},
\end{eqnarray}
 we now rather use a timelike Killing vector $\partial/\partial t$ and hence one should complete the square by $dt$  for a certain seed solution:
 \begin{eqnarray}
ds^2=-\hat\rho^2(dt+\hat B_{\hat \mu} dx^{\hat \mu})^2+\hat\rho^{-1}d\hat s^2_{(4)}.
\end{eqnarray}
Here $\mu,\nu,\cdots$ runs indexes except $x^5$ and $\hat\mu,\hat\nu,\cdots$ runs ones except $t$. $ds_{(4)}=g_{\mu\nu}^{(4)}dx^\mu dx^\nu$ and $d\hat s_{(4)}=\hat g_{\hat\mu\hat\nu}^{(4)}dx^{\hat\mu} dx^{\hat \nu}$ are the four-dimensional metrics on the dimensionally reduced spacetime and space, respectively.
%($\mu,\nu=t,r,\theta,\phi;
%\hat\mu,\hat\nu=r,\theta,\phi,x^5$)
In this paper, we call the operation {\it flip}.
In general, this operation changes a set of the fields $(g_{\mu\nu}^{(4)},B_\mu,A_\mu,\rho,A_5)$ to a set of different fields $(\hat g_{\hat\mu\hat\nu}^{(4)},\hat B_{\hat\mu},\hat A_{\hat\mu},\hat \rho, \hat A_t)$, 
while the flip itself does not 
%change the seed solution to another different 
generate any new solutions, {\it,i.e.}, the solution described by the fields $(\hat g_{\hat\mu\hat\nu}^{(4)},\hat B_{\hat\mu},\hat A_{\hat\mu},\hat \rho, \hat A_t)$  is the same as the one  %a set of the fields 
by $(g_{\mu\nu}^{(4)},B_\mu,A_\mu,\rho,A_5)$. 
After the flip, one performs the $SL(2,\RRsmall)$-duality transformation for the flipped seed solution and then can obtain, in general, a different solution in the bosonic sector of $D=5$ minimal supergravity (we denote these field by $new$). 
 Finally, one again flip the solution, {\it i.e.}, one completes the square by $dx^5$ rather than $dt$ for the obtained solution. Through this paper, we denote the flipped fields by attaching {\it hat}. 
To summarize, the procedure of obtaining new solutions by the series of transformations is as follows:  
\begin{eqnarray}
&&(g_{\mu\nu}^{(4)},B_\mu,A_\mu,\rho,A_5)\nonumber\\
&&\to{\it Flip}\nonumber
\to(\hat g_{\hat\mu\hat\nu}^{(4)},\hat B_{\hat\mu},\hat A_{\hat\mu},\hat \rho, \hat A_t)\\
&&\to{ SL(2,\RRsmall)}{\it \ duality\ transformation}\to(\hat g_{\hat\mu\hat\nu}^{(4)new},\hat B_{\hat\mu}^{new},\hat A_{\hat\mu}^{new},\hat \rho_{new},\hat A_t^{new})\nonumber\\
&&\to{\it Flip}\to(g_{\mu\nu}^{(4)new},B_{\mu}^{new},A_{\mu}^{new},\rho_{new},A_5^{new}).\nonumber
\end{eqnarray}
Further, though the below formula,  we assume that the spacetime also admits another spacelike Killing vector $\partial/\partial\phi$ which commutes with the other two Killing vectors. In general, this symmetry assumption is not necessary required for our $SL(2,\RRsmall)$-duality transformation. This is simply for later convenience and, actually, in the following section, we will apply our transformation to the Rasheed solutions which have this symmetry. 
In this case, it can be shown that a two-surface orthogonal to the three Killing vector fields are integrable~\cite{weyl,TYI}. For the integral two-surface, we use two coordinates $(r,\theta)$.

\subsection{Flip}
By completing the square by $dt$,
% for a standard metric form in the five-dimensional Kaluza-Klein theory, 
we can easily obtain the flipped scalar fields $(\hat \rho,\hat A_t)$, $U(1)$ gauge fields $(\hat B_{\hat \mu},\hat A_{\hat \mu})$ and four dimensional metric $\hat g_{\hat \mu\hat \nu}^{(4)}$. 
After the flip operation, the dilaton and axion fields can be written in the form:
\begin{eqnarray}
\hat\rho^2=-\left(\rho^2 B_t^2+\rho^{-1}g_{tt}^{(4)}\right),\label{eq:flip-rho}
\end{eqnarray}
\begin{eqnarray}
\hat A_t=A_t.\label{eq:flip-axion}
\end{eqnarray}
The gauge potential $1$-forms for the Kaluza-Klein $U(1)$ field and Maxwell $U(1)$ field are, respectively,
\begin{eqnarray}
\hat B_{\hat \mu}dx^{\hat \mu}=-\frac{\rho^2 B_tB_\phi+\rho^{-1}g_{t\phi}^{(4)}}{\hat\rho^{2}}d\phi-\frac{\rho^2 B_t}{\hat\rho^{2}} dx^5,\label{eq:flip-B}
\end{eqnarray}
\begin{eqnarray}
\hat A_{\hat \mu}=A_{\hat \mu}.\label{eq:flip-A}
\end{eqnarray}
%\begin{eqnarray}
%&&\hat B_\phi=\frac{\rho^2 B_tB_\phi+\rho^{-1}g_{t\phi}^{(4)}}{-\hat\rho^{2}},\\
%&&\hat B_5=\frac{\rho^2 B_t}{-\hat\rho^{2}},\\
%\end{eqnarray}
The four-dimensional metric is
\begin{eqnarray}
d\hat s^2_{(4)}&=&-\frac{\rho}{\hat\rho}\left[g_{tt}^{(4)}\left\{dx^5+\left(B_\phi-\frac{g_{t\phi}^{(4)}}{g_{tt}^{(4)}}B_t\right)d\phi\right\}^2-\frac{g_{\phi\phi}^{(4)}g_{tt}^{(4)}-g_{t\phi}^{(4)}{}^2}{g_{tt}^{(4)}} \frac{\hat\rho^2}{\rho^2}d\phi^2\right]\nonumber\\
&&+\frac{\hat \rho}{\rho}\left( g_{rr}^{(4)}dr^2+ g_{\theta\theta}^{(4)}d\theta^2\right).\label{eq:flip-metric}
\end{eqnarray}
%\begin{eqnarray}
%d\hat s^2_{(4)}&=&-\frac{\rho}{\hat\rho}\left[g_{tt}^{(4)}\left\{dx^5+\left(B_\phi-\frac{g_{t\phi}^{(4)}}{g_{tt}^{(4)}}B_t\right)d\phi\right\}^2+(g_{\phi\phi}^{(4)}g_{tt}^{(4)}-g_{t\phi}^{(4)}{}^2)\left(\frac{B_t^2}{g_{tt}^{(4)}}+\rho^{-3}\right)d\phi^2\right]\nonumber\\
%&&+\frac{\hat \rho}{\rho}\left( g_{rr}^{(4)}dr^2+ g_{\theta\theta}^{(4)}d\theta^2\right).\label{eq:flip-metric}
%\end{eqnarray}

\if0%%%%%%%%%
\begin{eqnarray}
g_{55}^{(4)'}=\frac{\rho^2+(\rho')^2(B_5')^2}{(\rho^{'})^{-1}}
\end{eqnarray}
\begin{eqnarray}
g_{5\phi}^{(4)'}=\frac{\rho^2B_\phi+(\rho')^2(B'_5)(B'_\phi)}{(\rho^{'})^{-1}}
\end{eqnarray}
\begin{eqnarray}
g_{\phi\phi}^{(4)'}=\frac{\rho^2B_\phi^2+\rho^{-1}g_{\phi\phi}^{(4)}+(\rho')^2(B'_\phi)^2}{(\rho^{'})^{-1}}
\end{eqnarray}
\begin{eqnarray}
g_{rr}^{(4)'}=\frac{\rho'}{\rho}g_{rr}^{(4)}
\end{eqnarray}
\begin{eqnarray}
g_{\theta\theta}^{(4)'}=\frac{\rho'}{\rho}g_{\theta\theta}^{(4)}
\end{eqnarray}
\fi

\subsection{SL(2,\RRsub)-duality transformation}
In general, performing the $SL(2,\RRsmall)$-duality transformation 
on the flipped fields (\ref{eq:flip-rho})-(\ref{eq:flip-metric}) yields a different solution from the 
one obtained by starting from the unflipped (hence, original) fields. 
%
%As shown in the previous section, after the transformation, 
%
%We set $\xi$ to be $-1$ and consider the $SL(2,\RRsmall)$ matrix (\ref{Lambda_gamma=0})
%999
%(\ref{Lambda})with $\gamma=1$, or
%\beqa
%\Lambda&=&\exp(-\delta E')\exp(-\epsilon F'),
%\eeqa
%where $\delta$ and $\epsilon$ are parameters. 
%999
%The possible Cartan-subalgebra factor 
%has been omitted as it does not add any new parameter to the solution \cite{MT}.
According to %(\ref{VVtoVVU-1})
(\ref{2Dscalartransformation}), the dilaton and axion fields for the new solution take the forms, respectively, 
\begin{eqnarray}
\hat\rho_{new}=\frac{\hat\rho}{\left\{1+\epsilon\left(\delta+\frac{\hat A_t}{\sqrt{3}}\right)\right\}^2-\epsilon^2\hat\rho^2},\label{eq:frhonew}
\end{eqnarray}
\begin{eqnarray}
\hat A_t^{new}=\sqrt{3}\frac{\left(1+\epsilon\delta+\epsilon\frac{\hat A_t}{\sqrt{3}}\right)\left(\delta+\frac{\hat A_t}{\sqrt{3}}\right)-\epsilon\hat\rho^2}{\left\{1+\epsilon\left(\delta+\frac{\hat A_t}{\sqrt{3}}\right)\right\}^2-\epsilon^2\hat\rho^2}.\label{eq:fAtnew}
\end{eqnarray}
Also, by (\ref{eq:Ftr}),
the two $U(1)$ gauge fields are transformed to 
\begin{eqnarray}
%\hat B_{\hat \mu}^{new}=(1+\epsilon\delta)^3\hat B_{\hat \mu}+\epsilon^3 \hat {\tilde B}_{\hat \mu}.\label{eq:fBnew}\\
{\hat B}^{new}_{\hat\mu} &= &
   \sqrt{3}  \epsilon^2(1+\delta\epsilon)\hat{\tilde A}_{\hat \mu}
   +(1+\delta\epsilon)^3 \hat B_{\hat \mu}
    -\sqrt{3}  \epsilon(1+\delta\epsilon)^2 \hat {A'}_{\hat \mu}
   +\epsilon^3 \hat {\cal H}_{\hat\mu}^{B},\label{eq:fBnew}\\
%
%\hat A_{\hat \mu}^{new}=\left[(1+\epsilon\delta)^3\hat A_t^{new}-\sqrt{3}\delta  (1+\epsilon\delta)^2\right] \hat B_{\hat \mu}+\left(\epsilon^3\hat A_t^{new}-\sqrt{3}\epsilon^2\right)\hat {\tilde B}_{\hat \mu}.\label{eq:fAnew}
%\text{A$\phi $pnew}=\left.(\zeta {}^{\wedge}3\text{Atpnew}-\sqrt{3}\delta  \zeta {}^{\wedge}2\right)\text{B$\phi $p}+\left(\epsilon {}^{\wedge}3\text{Atpnew}-\sqrt{3}\epsilon {}^{\wedge}2)\right.\text{tB$\phi $p}
\hat {A'}^{new}_{\hat\mu} &= &
-\left(3 \delta  \epsilon ^2+2 \epsilon
   \right)\hat{\tilde A}_{\hat \mu}- \sqrt{3} \delta(1+\delta\epsilon)^2\hat B_{\hat \mu}  
   +\left(3 \delta ^2 \epsilon ^2+4 \delta  \epsilon
   +1\right) \hat {A'}_{\hat \mu}
   -\sqrt{3}  \epsilon ^2 \hat {\cal H}_{\hat\mu}^{B}.\label{eq:fAnew}
\end{eqnarray}
%
%
%\begin{eqnarray}
%B_{\mu}^{'new}=\sqrt{3}\epsilon^2\zeta \tilde A_{\mu}'+\zeta^3B_{\mu}'-\sqrt{3}\epsilon \zeta^2(A_{\mu}')'+\epsilon^3 {\cal H}_{\mu}^{B\prime}
%\end{eqnarray}
%\begin{eqnarray}
%\hat\rho^{new}=\frac{\hat\rho}{\left\{1+\delta\left(\epsilon+\frac{\hat A_t}{\sqrt{3}}\right)\right\}^2-\delta^2\hat\rho^2},
%\end{eqnarray}
%\begin{eqnarray}
%-(A_{\mu}')^{'new}=\epsilon(2+3\epsilon\delta) \tilde A_{\mu}'+\sqrt{3}\epsilon\zeta^2B_{\mu}'-(1+4\epsilon\delta+3\epsilon^2 \delta^2)(A_{\mu}')^{'}+\sqrt{3}\epsilon^2 {\cal H}_{\mu}^{B'}
%\end{eqnarray}
Under the transformation, the four-dimensional metric is invariant:
\begin{eqnarray}
\hat g_{\hat\mu\hat\nu}^{(4)new}=\hat g_{\hat\mu\hat\nu}^{(4)}.
\end{eqnarray}

\subsection{Flip again }\label{sec:2flip}
Finally, in order to write the fields $(\hat g_{\hat\mu\hat\nu}^{(4)new},\hat B_{\hat\mu}^{new},\hat A_{\hat\mu}^{new},\hat \rho_{new},\hat A_t^{new})$ in the standard Kaluza-Klein form (though this is not always required), we again must flip the new solution by completing the square by $dx^5$. 
After performing the second flip, the dilaton and axion field $(\rho_{new},A_5^{new})$ for the new solution are in the following forms, respectively,
\begin{eqnarray}
\rho_{new}^2=-(\hat\rho_{new})^2(\hat B_5^{new})^2+(\hat\rho_{new})^{-1}\hat g_{55}^{new},
\end{eqnarray}
\begin{eqnarray}
A_5^{new}=\hat A_5^{new}.
\end{eqnarray}
The two $U(1)$ gauge fields are
\begin{eqnarray}
B^{new}_{\mu}dx^\mu=-\frac{(\hat\rho_{new})^2(\hat B_5^{new})}{\rho^2_{new}} dt-\frac{(\hat\rho_{new})^2(\hat B_5^{new})(\hat B_\phi^{new})+(\hat \rho_{new})^{-1}\hat g_{5\phi}^{(4)new}}{\rho^2_{new}}d\phi,
\end{eqnarray}
\begin{eqnarray}
A^{new}_{\mu}=\hat A_\mu^{new}.
\end{eqnarray}
%\begin{eqnarray}
%B_\phi^{new}=\frac{-(\rho'_{new})^2(B_5^{'new})(B_\phi^{'new})+(\rho'_{new})^{-1}g_{5\phi}^{(4)'new}}{\rho^2_{new}}
%\end{eqnarray}
%\begin{eqnarray}
%B_t^{new}=\frac{-(\rho'_{new})^2(B_5^{'new})}{\rho^2_{new}}
%\end{eqnarray}
The dimensionally reduced four-dimensional metric $ds^2_{(4)new}=g_{\mu\nu}^{(4)new}dx^\mu dx^\nu$ is given by
\begin{eqnarray}
ds^2_{(4)new}&=&-\frac{\hat\rho_{new}}{\rho_{new}}\Biggl[\hat g_{55}^{(4)new}\left\{dt+\left(\hat B_\phi-\frac{\hat g_{5\phi}^{(4)new}}{\hat g_{55}^{(4)new}}\hat B_5^{new}\right)d\phi\right\}^2
{}-
\frac{\hat g_{\phi\phi}^{(4)new}\hat g_{55}^{(4)new}-{\hat g_{5\phi}^{(4)new}{}^2}}{\hat g_{55}^{(4)new}}\frac{\rho_{new}^2}{ \hat\rho_{new}^2}d\phi^2\Biggr]\nonumber\\ 
&&{ }+\frac{\rho_{new}}{\hat\rho_{new}}\left(\hat g_{rr}^{(4)new}dr^2+\hat g_{\theta\theta}^{(4)new}d\theta^2\right).\label{eq:4metricff}
\end{eqnarray}

\section{Applications}
\subsection{Rasheed solution}

In the following subsection, using the $SL(2,\RRsmall)$-transformation mentioned in the previous section,
we will generate a rotating black hole solution, starting from the Rasheed solution~\cite{Rasheed}.
Hence, in this section, we briefly review the Rasheed solutions in five-dimensional pure gravity ( 
We also apply our technique to 
asymptotically flat black hole solutions such as the five-dimensional Myers-Perry solutions~\cite{MP}. See Appendix A for this.).
The metric of the Rasheed solution is given by
\begin{eqnarray}
ds^2&=&\frac{B}{A}(dx^5+
B_\mu dx^\mu)^2+\sqrt{\frac{A}{B}}ds^2_{(4)},\label{eq:Rasheed}
\end{eqnarray}
where the four-dimensional (dimensionally reduced) metric is given by
\begin{eqnarray}
ds^2_{(4)}=-\frac{f^2}{\sqrt{AB}}(dt+\omega^0{}_\phi d\phi)^2+\frac{\sqrt{AB}}{\Delta}dr^2+\sqrt{AB}d\theta^2+\frac{\sqrt{AB}\Delta}{f^2}\sin^2\theta d\phi^2.\label{eq:4metric}
\end{eqnarray}
Here the functions $(A,B,C,\omega^0{}_\phi,\omega^5{}_{\phi},f^2,\Delta)$ and the $1$-form 
$B_{\mu}$ 
\footnote{To avoid confusion with the Maxwell field, we denote in this paper 
the Kaluza-Klein vector field as $B_{\mu}$,  instead of $2 A_\mu^{(Rasheed)}$ 
used in the original article \cite{Rasheed}.}
are
\begin{eqnarray}
&&A=\left(r-\frac{\Sigma}{\sqrt{3}}\right)^2-\frac{2P^2\Sigma}{\Sigma-\sqrt{3}M}+a^2\cos^2\theta+\frac{2JPQ\cos\theta}{(M+\Sigma/\sqrt{3})^2-Q^2},\\
&&B=\left(r+\frac{\Sigma}{\sqrt{3}}\right)^2-\frac{2Q^2\Sigma}{\Sigma+\sqrt{3}M}+a^2\cos^2\theta-\frac{2JPQ\cos\theta}{(M-\Sigma/\sqrt{3})^2-P^2},\\
&&C=2Q(r-\Sigma/\sqrt{3})-\frac{2PJ\cos\theta(M+\Sigma/\sqrt{3})}{(M-\Sigma/\sqrt{3})^2-P^2},\\
&&\omega^0{}_\phi=\frac{2J\sin^2\theta}{f^2}\left[r-M+\frac{(M^2+\Sigma^2-P^2-Q^2)(M+\Sigma/\sqrt{3})}{(M+\Sigma/\sqrt{3})^2-Q^2}\right],\\
&&\omega^5{}_\phi=\frac{2P\Delta \cos\theta}{f^2}-\frac{2QJ\sin^2\theta[r(M-\Sigma/\sqrt{3})+M\Sigma/\sqrt{3}+\Sigma^2-P^2-Q^2]}{f^2[(M+\Sigma/\sqrt{3})^2-Q^2]},\\
&&\Delta=r^2-2Mr+P^2+Q^2-\Sigma^2+a^2,\\
&&f^2=r^2-2Mr +P^2+Q^2-\Sigma^2+a^2\cos^2\theta,\\
&& 
B_\mu dx^\mu=\frac{C}{B}dt+\left(\omega^5{}_\phi+\frac{C}{B}\omega^0{}_\phi\right)d\phi,\label{eq:RasheedB}
\end{eqnarray}
where $B_\mu$ describes the electromagnetic vector potential derived by dimensional reduction to four dimension. Here the constants, $(M,P,Q,J,\Sigma)$, mean the mass, Kaluza-Klein magnetic charge, Kaluza-Klein electric charge, angular momentum along four dimension and dilaton charge, respectively, which are parameterized by the two parameters $(\hat\alpha,\hat\beta)$
\begin{eqnarray}
&&M=\frac{(1+\cosh^2\hat\alpha\cosh^2\hat\beta)\cosh\hat\alpha}{2\sqrt{1+\sinh^2\hat\alpha\cosh^2\hat\beta}}M_k,\\
&&\Sigma=\frac{\sqrt{3}\cosh\hat\alpha(1-\cosh^2\hat\beta+\sinh^2\hat\alpha\cosh^2\hat\beta)}{2\sqrt{1+\sinh^2\hat\alpha\cosh^2\hat\beta}}M_k,\\
&&Q=\sinh\hat\alpha\sqrt{1+\sinh^2\hat\alpha\cosh^2\hat\beta}\ M_k,\\
&&P=\frac{\sinh\hat\beta \cosh\hat\beta}{\sqrt{1+\sinh^2\hat\alpha\cosh^2\hat\beta}}M_k,\\
&&J= \cosh\hat\beta \sqrt{1+\sinh^2\hat\alpha\cosh^2\hat\beta\ }aM_k.
\end{eqnarray}
Note that all the above parameters are not independent since they are related through the equation
\begin{eqnarray}
\frac{Q^2}{\Sigma+\sqrt{3}M}+\frac{P^2}{\Sigma-\sqrt{3}M}=\frac{2\Sigma}{3},\label{eq:constraint}
\end{eqnarray}
and the constant $M_k$ is written in terms of these parameters
\begin{eqnarray}
M_k^2=M^2+\Sigma^2-P^2-Q^2.
\end{eqnarray}
The constant $J$ is also related to $a$ by
\begin{eqnarray}
J^2=a^2\frac{[(M+\Sigma/\sqrt{3})^2-Q^2][(M-\Sigma/\sqrt{3})^2-P^2]}{M^2+\Sigma^2-P^2-Q^2}.
\end{eqnarray}

\medskip
The dilaton and axion fields for the Rasheed solution are, respectively,
\begin{eqnarray}
\rho=\sqrt{\frac{B}{A}},\quad A_5=0.\label{eq:rhonew}
\end{eqnarray}
The Kaluza-Klein gauge field and Maxwell field are, respectively,
\begin{eqnarray}
B_\mu dx^\mu=\frac{C}{B}dt+\left(\omega^5{}_\phi+\frac{C}{B}\omega^0{}_\phi\right)d\phi,\quad 
A_\mu dx^\mu=0.
\end{eqnarray}

\subsection{Flipped Rasheed solution}
By completing the square for the time coordinate $t$, the metric of the flipped Rasheed solution is obtained:
\begin{eqnarray}
ds^2&=&-\frac{Af^2-C^2}{AB}(dt+
\hat B_{\hat \mu} dx^{\hat \mu})^2+\sqrt{\frac{AB}{Af^2-C^2}}d\hat s^2_{(4)},\label{eq:Flipped Rasheed}
\end{eqnarray}
where the gauge potential $1$-form for Kaluza-Klein $U(1)$ gauge field is
\begin{eqnarray}
\hat B_{\hat\mu}dx^{\hat\mu}=\frac{-BC}{Af^2-C^2}dx^5+\left(\frac{-BC}{Af^2-C^2}\omega^5{}_\phi+\omega^0{}_\phi\right)d\phi,
\end{eqnarray}
%\begin{eqnarray}
%&&\tilde B_5=\frac{-BC}{Af^2-C^2},\\
%&&\tilde B_\phi=\frac{-BC}{Af^2-C^2}\omega^5{}_\phi+\omega^0{}_\phi,\\
%\end{eqnarray}
and the four-dimensional metric is
\begin{eqnarray}
d\hat s^2_{(4)}=\hat\rho\left[\frac{Bf^2}{Af^2-C^2}\left(dx^5+\omega^5{}_\phi d\phi\right)^2+\frac{A\Delta}{f^2}\sin^2\theta d\phi^2+A\left(\frac{dr^2}{\Delta}+d\theta^2\right)\right].
\end{eqnarray}
\begin{eqnarray}
\end{eqnarray}
From this, the dilaton and axion read
\begin{eqnarray}
\hat\rho=\sqrt{\frac{Af^2-C^2}{AB}},
\end{eqnarray}
\begin{eqnarray}
\hat A_t=0.
\end{eqnarray}
The gauge potential one-form for Maxwell field are
\begin{eqnarray}
\hat A_{\hat\mu}=0.
\end{eqnarray}

\subsection{Transformed Rasheed solutions}
Applying the $SL(2,\RRsmall)$-duality transformation to the flipped Rasheed solution, we can obtain new Kaluza-Klein black hole solutions in $D=5$ minimal supergravity.
By putting $\hat A_t=0$ in Eqs. (\ref{eq:frhonew}) and (\ref{eq:fAtnew}), the two scalar fields for new solutions are written as
\begin{eqnarray}
\hat\rho_{new}=\frac{\hat\rho}{(1+\epsilon\delta)^2-\epsilon^2\hat\rho^2},
\end{eqnarray}
\begin{eqnarray}
\hat A_t^{new}=\sqrt{3}\frac{\left(1+\epsilon\delta\right)\delta -\epsilon\hat\rho^2}{(1+\epsilon\delta)^2-\epsilon^2\hat\rho^2}.
\end{eqnarray}

Putting $\hat {\tilde A}_{\hat \mu}=\hat A'_{\hat \mu}=0$ in Eqs.~(\ref{eq:fBnew}) and (\ref{eq:fAnew}), we immediately obtain the two vector fields as
\begin{eqnarray}
B^{new}_{\hat\mu}=(1+\delta\epsilon)^3B_{\hat\mu}+\epsilon^3\hat{\cal H}^B_{\hat\mu},
\end{eqnarray}
\begin{eqnarray}
\hat A_{\hat\mu}^{new}=\left[(1+\delta\epsilon)^3\hat A_t^{new}-\sqrt{3}\delta  (1+\delta\epsilon)^2\right]B_{\hat\mu}
+\left(\epsilon^3\hat A_t^{new}-\sqrt{3}\epsilon^2\right)\hat{\cal H}^B_{\hat\mu},
\end{eqnarray}
where $\hat {\cal H}^B_{\hat \mu}$ can be obtain by actually integrating Eq.~(\ref{eq:HB}) as
\begin{eqnarray}
\hat {\cal H}^B_5=\frac{-2 P Q-2 J \cos\theta}{A},
\end{eqnarray}
\begin{eqnarray}
\hat {\cal H}^B_\phi=\left(b_0 r+\frac{\left({c_2} r^2+{c_1} r+{c_0}\right)\cos\theta+{d_3} r^3+{d_2} r^2+{d_1} r+{d_0}}{A}\right),
\end{eqnarray}
where the constants $b_0,c_0,c_1,c_2,d_0,d_1,d_2,d_3$ are defined by:
\begin{eqnarray}
b_0=-d_3=\frac{2 J P \left(M+ \Sigma/\sqrt{3} \right)}{a^2 \left[\left(M-\Sigma/\sqrt{3}\right)^2- P^2\right]},
\end{eqnarray}
\begin{eqnarray}
c_0&=&\frac{1}{2} Q \biggl[-4 a^2-\frac{4 \Sigma ^2}{3}+\frac{8 \sqrt{3} M P^2}{-\sqrt{3} M+\Sigma }\nonumber\\
&&+(24 J^2 P^2 (54 M^6+63 \sqrt{3} M^5 \Sigma -9 M^4 (6 P^2+12 Q^2-13 \Sigma ^2)\nonumber\\
&&+6 \sqrt{3} M^3 \Sigma  (-12 P^2-15 Q^2+11 \Sigma ^2)-(3 Q^2-\Sigma ^2) (9 Q^2 (P^2+Q^2)-3 (2 P^2+5 Q^2) \Sigma ^2+5 \Sigma ^4)\nonumber\\
&&+3 M^2 (27 Q^2 (P^2+Q^2)-18 (2 P^2+3 Q^2) \Sigma ^2+28 \Sigma ^4)+\sqrt{3} M \Sigma  (45 Q^4-66 Q^2 \Sigma ^2+19 \Sigma ^4+6 P^2 (9 Q^2-4 \Sigma ^2))))\nonumber\\
&&\times\left[81a^2 \left[\left(M-\Sigma/\sqrt{3}\right)^2- P^2\right] \left[\left(M+\Sigma /\sqrt{3}\right)^2-Q^2\right]^3\right]^{-1}\biggr],
\end{eqnarray}
\begin{eqnarray}
&&c_1=\frac{4 Q \Sigma }{\sqrt{3}}-\frac{4 J^2 P^2 Q \left(M+\Sigma/\sqrt{3}  \right)}{a^2 \left[\left(M-\Sigma/\sqrt{3}\right)^2- P^2\right] \left[\left(M+\Sigma /\sqrt{3}\right)^2-Q^2\right]},\\
&&c_2=-2 Q,
\end{eqnarray}
\begin{eqnarray}
d_0&=&\biggl[2 J P (81 \sqrt{3} a^2 M^2 Q^2 (-2 M^2+P^2+Q^2)+81 M (a^2 (3 M^4+M^2 Q^2-2 Q^2 (P^2+Q^2))\nonumber\\
&&+2 P^2 (2 M^4-2 M^2 (P^2+Q^2)+Q^2 (P^2+Q^2))) \Sigma +27 \sqrt{3} (3 a^2 M^4+2 M^6-4 M^2 P^4\nonumber\\
&&-(2 M^4+M^2 P^2+2 P^4+a^2 (M-P) (M+P)) Q^2\nonumber\\
&&+(a^2+M^2-2 P^2) Q^4) \Sigma ^2-27 M (6 a^2 M^2+M^4-4 M^2 P^2-4 P^4-(7 a^2+M^2-6 P^2) Q^2+2 Q^4) \Sigma ^3\nonumber\\
&&+9 \sqrt{3} (-6 a^2 M^2+M^4+12 M^2 P^2+4 P^4-(5 a^2+M^2-11 P^2) Q^2+Q^4) \Sigma ^4+9 M (3 a^2+2 M^2-8 P^2+7 Q^2) \Sigma ^5\nonumber\\
&&-3 \sqrt{3} (-3 a^2+8 M^2+12 P^2+5 Q^2) \Sigma ^6-3 M \Sigma ^7+5 \sqrt{3} \Sigma ^8)\biggr]\nonumber\\
&&\times\left[81\sqrt{3} a^2 ( M-\Sigma/\sqrt{3} )^2\left[\left(M-\Sigma/\sqrt{3}\right)^2- P^2\right] \left[\left(M+\Sigma /\sqrt{3}\right)^2-Q^2\right]\right]^{-1}
\end{eqnarray}

\begin{eqnarray}
d_1&=&\biggl[2 J P (27 \sqrt{3} a^2 M^2 (M-Q) (M+Q)+54 M (a^2 M^2+2 M^4+Q^4-M^2 (P^2+2 Q^2)) \Sigma \nonumber\\
&&+9 \sqrt{3} (3 M^4+M^2 (2 P^2-3 Q^2)+Q^2 (a^2-2 (2 P^2+Q^2))) \Sigma ^2-18 M (a^2-3 M^2-5 P^2+2 Q^2) \Sigma ^3\nonumber\\
&&+3 \sqrt{3} (-a^2+8 M^2+6 P^2+11 Q^2) \Sigma ^4-30 M \Sigma ^5-11 \sqrt{3} \Sigma ^6)\biggr]\nonumber\\
&&\times\left[-27\sqrt{3} a^2 (M-\Sigma/\sqrt{3})\left[\left(M-\Sigma/\sqrt{3}\right)^2- P^2\right] \left[\left(M+\Sigma /\sqrt{3}\right)^2-Q^2\right] \right]^{-1},
\end{eqnarray}
\begin{eqnarray}
d_2&=&-\biggl[2 J P (-18 M^4-9 Q^2 (P^2+Q^2)-15 \sqrt{3} M^3 \Sigma +3 (2 P^2+7 Q^2) \Sigma ^2-7 \Sigma ^4+3 M^2 (6 (P^2+Q^2)-11 \Sigma ^2)\nonumber\\
&&+3 \sqrt{3} M \Sigma  (4 P^2+5 (Q-\Sigma ) (Q+\Sigma )))\biggr]\left[9a^2\left[\left(M-\Sigma/\sqrt{3}\right)^2- P^2\right] \left[\left(M+\Sigma /\sqrt{3}\right)^2-Q^2\right]\right]^{-1}.
\end{eqnarray}

\subsection{Flipped transformed Rasheed solution}
 By flipping the metric and Maxwell's $U(1)$ field again according to Sec.~\ref{sec:2flip}, one can read off the (Lorentzian) metric and Kaluza-Klein's $U(1)$ field and Maxwell's $U(1)$ field for the four-dimensional dimensionally reduced  spacetime.

\subsection{Asymptotics}
Now we investigate asymptotics of the obtained solution. 
It turns out that the solutions do not have Kaluza-Klein asymptotics in a usual sense 
since after the flip  the $t\phi$-component of the dimensionally reduced four-dimensional metric (\ref{eq:4metricff}) at infinity $r\to\infty$ behaves as
\begin{eqnarray}
g_{t\phi}^{(4)new}\simeq \frac{1}{N^{3/2}}\left(c+2Q\epsilon^2\cos\theta\right)+{\cal O}(r^{-1}),
\end{eqnarray}
where the constants $N$ and $c_\phi$ are
\begin{eqnarray}
N=(1+\delta\epsilon)^2-\epsilon^2,
\end{eqnarray}
\begin{eqnarray}
c&=&-\Biggl[2 J P \epsilon ^3 \left(18 M^4+9 Q^2 \left(P^2+Q^2\right)+9 \sqrt{3} M^3 \Sigma -3 \left(2 P^2+5 Q^2\right) \Sigma ^2+5 \Sigma ^4-3 M^2 \biggl(6 \left(P^2+Q^2\right)-5 \Sigma ^2\right)\nonumber\\
&&+3 \sqrt{3} M \Sigma  \left(-4 P^2-3 Q^2+3 \Sigma ^2\right)\biggr)\Biggr]\nonumber\\
&&\times \left[a^2  \left(3 M^2-3 P^2-2 \sqrt{3} M \Sigma +\Sigma ^2\right) \left(3 M^2-3 Q^2+2 \sqrt{3} M \Sigma +\Sigma ^2\right)\right]^{-1}.
\end{eqnarray}
The constant term vanishes under the coordinate transformation $t-c\to t$  but the term proportional to $\cos\theta$ does not vanishes (hence this is not a simply gauge).
The existence of this term means that the dimensionally reduced spacetime has the so-called NUT parameter. 
Though as a result, for our solutions the dimensionally reduced four-dimensional spacetime are generally not asymptotically Minkowskian,  
 if and only if $Q \epsilon=0$, this NUT parameter vanishes and our solutions describe usual Kaluza-Klein black holes.

%\begin{eqnarray}
%&&A_\phi'=A_\phi,\\
%&&A_t'=A_5,
%\end{eqnarray}

%\begin{eqnarray}
%ds^2_{(4)}{}'&=&\left(\rho'\rho^2-\rho^4B_t^2\rho'^{-1}\right)(dx^5)^2+2\left[\rho^2\rho'B_\phi-(\rho^2B_t B_\phi+\rho^{-1}g_{t\phi}^{(4)})\rho^2\rho'^{-1}B_t\right]d\phi dx^5\nonumber\\
%             &&\left[\rho'\rho^2B_\phi^2+\rho^{-1}\rho'g_{\phi\phi}^{(4)}-(\rho^2B_tB_\phi+\rho^{-1}g_{t\phi}^{(4)})^2\rho'^{-1}\right]d\phi^2+\frac{\rho'}{\rho}(g_{rr}^{(4)}dr^2+g_{\theta\theta}^{(4)}d\theta^2)
%\end{eqnarray}

\section{summary and discussion}
In this paper, in the choice of a timelike Killing vector, we have performed dimensional reduction to a four-dimensional Euclidean space and have also shown in that case the field equations are invariant under $SL(2,\RRsmall)/SO(1,1)$ transformation. 
 In the timelike case, we have also developed a new solution-generation technique using the duality transformation, as  we have done in~\cite{MT} for the spacelike case. As an example, by applying this transformation to the Rasheed solutions, we have obtained rotating Kaluza-Klein black hole solutions in five-dimensional minimal supergravity. In general, in contrast to the spacelike cases, the resulting dimensionally reduced solution includes the so-called NUT parameter and for this reason, in general, the dimensionally reduced spacetime is not asymptotically flat. However,  In some special cases (for instance, when the electric charge $Q$ for the Kaluza-Klein $U(1)$ field vanishes, it can describe ordinary Kaluza-Klein black holes.

\section*{Acknowledgments} 
The work of S.~M. and S.~T. is supported by 
Grant-in-Aid
for Scientific Research  
(A) No.22244030, and S.~M. 
is also by (C) No.20540287  
from
The Ministry of Education, Culture, Sports, Science
and Technology of Japan.

\appendix
\section{Myers-Perry black holes}
The flipped metric of the five-dimensional Myers-Perry solutions is given by
\begin{eqnarray}
ds^2&=&-\left(1-\frac{r_0^2}{\varrho^2}\right)\left(dt-\frac{r_0^2a}{\varrho^2-r_0^2}\sin^2\theta d\phi-\frac{r_0^2b}{\varrho^2-r_0^2}\cos^2\theta d\psi\right)^2+\left(x+a^2+\frac{r_0^2a^2}{\varrho^2-r_0^2}\sin^2\theta\right )\sin^2\theta d\phi^2\nonumber\\
&&+\left(x+b^2+\frac{r_0^2b}{\varrho-r_0^2}\cos^2\theta\right)\cos^2\theta+2\frac{r_0^2ab}{\varrho^2-r_0^2}\sin^2\theta\cos^2\theta d\phi d\psi+\frac{\varrho^2}{4\Delta}dx^2+\varrho^2d\theta^2,
\end{eqnarray}
where 
\begin{eqnarray}
\varrho^2=x+a^2\cos^2\theta+b^2\sin^2\theta,\ \Delta=(x+a^2)(x+b^2)-r_0^2x.
\end{eqnarray}
The dilaton and axion are, respectively,
\begin{eqnarray}
\hat\rho=\sqrt{1-\frac{r_0^2}{\varrho^2}},\ \hat A_t=0, 
\end{eqnarray}
and the Kaluza-Klein's and Maxwell's $U(1)$ fields are
\begin{eqnarray}
\hat B_{\hat \mu}dx^{\hat \mu}=-\frac{r_0^2a}{\varrho^2-r_0^2}\sin^2\theta d\phi-\frac{r_0^2b}{\varrho^2-r_0^2}\cos^2\theta d\psi,\ \hat A_{\hat \mu}dx^{\hat \mu}=0.
\end{eqnarray}
Integrating Eq.(\ref{eq:HB}), we obtain
\begin{eqnarray}
\hat H^{B}_{\hat\mu} dx^{\hat \mu}=\frac{br_0^2(x+a^2)}{(a^2-b^2)\varrho^2}d\phi-\frac{ar_0^2(x+b^2)}{(a^2-b^2)\varrho^2}d\psi
\end{eqnarray}
Consider the rescale of the coordinates:
\begin{eqnarray}
\frac{dt}{N}\to dt,\ Ndx\to dx,
\end{eqnarray}
the redefinition of the parameters:
\begin{eqnarray}
N^{1/2}r_0\to r_0,\ N^{1/2}a\to a,\ N^{1/2}b\to b,
\end{eqnarray}
where we have defined $N\equiv \gamma^2-\epsilon^2$ $(\gamma=1+\delta\epsilon)$. Further, transform the coordinates
\begin{eqnarray}
dt +\tilde\epsilon^3\frac{r_0^2b}{a^2-b^2}d\phi-\tilde\epsilon^3\frac{r_0^2a}{a^2-b^2}d\psi\to dt,
\end{eqnarray} 
where $\tilde\epsilon \equiv \epsilon/N^{1/2}$ and $\tilde \gamma=\gamma/N^{1/2}$.
After the redefinition and coordinate transformations, the metric (derived by the $SL(2,R)$-duality transformation) takes the following form:
\begin{eqnarray}
ds^2&=&-\frac{1}{\left[\tilde\gamma^2-\tilde \epsilon^2\left(1-r_0^2/\varrho^2\right)\right]^2}  \left(1-\frac{r_0^2}{\varrho^2}\right)\left[dt-r_0^2\left(\frac{\tilde\gamma^3a}{\varrho^2-r_0^2}-\frac{\tilde\epsilon^3b}{\varrho^2}\right)\sin^2\theta d\phi-r_0^2\left(\frac{\tilde\gamma^3b}{\varrho^2-r_0^2}-\frac{\tilde\epsilon^3a}{\varrho^2}\right)\cos^2\theta d\psi\right]^2\nonumber\\
&&+\left[\tilde\gamma^2-\tilde \epsilon^2\left(1-r_0^2/\varrho^2\right)\right]\biggl[\left(x+a^2+\frac{r_0^2a^2}{\varrho^2-r_0^2}\sin^2\theta\right )\sin^2\theta d\phi^2\nonumber\\
&&+\left(x+b^2+\frac{r_0^2b}{\varrho-r_0^2}\cos^2\theta\right)\cos^2\theta+2\frac{r_0^2ab}{\varrho^2-r_0^2}\sin^2\theta\cos^2\theta d\phi d\psi+\frac{\varrho^2}{4}dx^2+\varrho^2d\theta^2\biggr].
\end{eqnarray}
This coincides with the metric form of the Cveti{\v c}-Youm solution~\cite{CY96} which was (re)derived by the $G_{2(+2)}$ transformation in Ref.~\cite{BCCGSW} (Note that 
$\tilde \gamma =c$ and $\tilde\epsilon=-s$). See Appendix B for the relationship 
between the two, $SL(2,R)$ and $G_{2(+2)}$, transformations.

%\appendix
\section{Relation to the Harrison transformation in the $G_{2(+2)}$ duality}
In this Appendix we clarify how the time-like $SL(2,R)$ transformation, investigated 
in this paper, is embedded into $G_{2(+2)}$ if the given seed solution allows another 
spacelike Killing vector $\frac\partial{\partial z}$. 
A similar analysis for the case when the two Killing vectors are both spacelike 
was already done in \cite{Germar}.

We decompose the four-dimensional 
metric (vielbein) and gauge field as
\beqa
E^{(4)\alpha}_{~\mu}&=&\left(
\begin{array}{cc}
e^{\phi}E^{(3)a}_{~m}
&C_m e^{-\phi} \\
0& e^{-\phi}
\end{array}
\right),\\
A_\mu&=&(A_m, A_z),
\eeqa
and write
\beqa
E^{(5)A}_{~M}&=&\left(
\begin{array}{cc}
e^{-1} E^{(3)a}_{~m}
&B_m^i e_i^{\bar a} \\
0& e_i^{\bar a}
\end{array}
\right),\\
A_M&=&(A_m, A_i),
\eeqa
\beqa
\eta_{AB}=\left(
\begin{array}{cc}
\delta_{ab}
&0 \\
0& \eta_{\bar{a}\bar{b}}
\end{array}
\right),~~
\eta_{\bar{a}\bar{b}}=\mbox{diag}(+1,-1),
\eeqa
\beqa
e_i^{\bar a}=\left(
\begin{array}{cc}
\rho^{-\frac12}e^{-\phi}&\rho B_z\\
0&\rho
\end{array}
\right),~~e=\mbox{det}e_i^{\bar a},~~B_m^i=(C_m,B_m).
\eeqa
Here
$m$ ($a$) is the three-dimensional curved (flat) index,
$i=t,z$ and $\bar a=1,2$.
All the fields are assumed to be independent of $t$ and $z$.
Then the reduced Lagrangian reads 
\beqa
{\cal L}&=&E^{(3)}\left(
R^{(3)}+\frac14\partial_m g^{ij} \partial^m g_{ij}
-e^{-2}\partial_m e \partial^m e
-\frac12 g^{ij} \partial_m A_i \partial^m A_j\right.\n
&&\left.
-\frac14 e^2 g_{ij} B_{mn}^i B^{jmn}
-\frac14 e^2 F^{(3)}_{mn} F^{(3)mn}
-\frac1{2\sqrt{3}}E^{(3)-1}\epsilon^{mnp}\epsilon^{ij}F_{mn}\partial_p A_i ~ A_j
\right),
\eeqa
where
$g_{ij}=e_i^{\bar a}\eta_{\bar{a}\bar{b}}e_j^{\bar b}$ and $F^{(3)}_{mn}=F_{mn}-2B_{[n}^i \partial_{m]}A_i$.
The vector fields $B_m^i$ and $A_m$ are dualized by adding the Lagrange multiplier terms
\beqa
{\cal L}^{5\rightarrow 3}_{Lag.mult.}&=&
-\frac12\epsilon^{mnp}\left(F'_{mn} \partial_p \varphi +
B^i_{mn}\partial_p \psi_i 
\right)
\eeqa
and completing the squares. Up to duality relations we have
\beqa
{\cal L}+{\cal L}^{5\rightarrow 3}_{Lag.mult.}
&=&E^{(3)}\left(
R^{(3)}+\frac14\partial_m g^{ij} \partial^m g_{ij}
-e^{-2}\partial_m e \partial^m e
-\frac12 g^{ij} \partial_m A_i \partial^m A_j\right.\n
&&
+\frac12 e^{-2}(\partial_m\varphi-\frac1{\sqrt{3}}\epsilon^{ij}A_i\partial_m A_j)
(\partial^m\varphi-\frac1{\sqrt{3}}\epsilon^{kl}A_k\partial^m A_l),
\n
&&\left.
+\frac12 e^{-2}g^{ij}(\partial_m\psi_i-A_i\partial_m\varphi+
\frac1{3\sqrt{3}}\epsilon^{kl}A_iA_k\partial_m A_l)
(\partial^m\psi_j-A_j\partial^m\varphi+
\frac1{3\sqrt{3}}\epsilon^{k'l'}A_jA_{k'}\partial^m A_{l'})
\right).\n
\label{3DscalarLagrangian}
\eeqa
The duality relations are
\beqa
F^{(3)}_{mn}&=&-e^{-2}E^{(3)-1}\epsilon_{mn}^{~~~~p}
(\partial_p\varphi-\frac1{\sqrt{3}}\epsilon^{ij}A_i\partial_p A_j),\n
B^i_{mn}&=&-e^{-2}E^{(3)-1}\epsilon_{mn}^{~~~~p}
(\partial_p\psi_i-A_i\partial_p\varphi+
\frac1{3\sqrt{3}}\epsilon^{kl}A_iA_k\partial_p A_l).
\eeqa
From (\ref{3DscalarLagrangian}) the target space metric can be read off, 
in terms of a matrix $({\bf g})_{ij}=g_{ij}$ and vectors $(\vec{A})_i=A_i$ and 
$(\vec{\psi})_i=\psi_i$, 
as
\beqa
ds^2_{target}&=&\frac14 \mbox{Tr}({\bf g}^{-1}d{\bf g})^2+e^{-2} de^2
+\frac12 d\vec{A}^T {\bf g}^{-1} d\vec{A}
-\frac12 e^{-2}\left(d\varphi-\frac1{\sqrt{3}}\epsilon^{ij}A_idA_j\right)^2\n
&&-\frac12 e^{-2}\left(
d\vec{\psi}-\vec{A}\left(d\varphi-\frac1{3\sqrt{3}}\epsilon^{ij}A_idA_j)\right)^T
{\bf g}^{-1}\left(d\vec{\psi}-\vec{A}(d\varphi-\frac1{3\sqrt{3}}\epsilon^{kl}A_kdA_l\right)
\right).
\label{target}
\eeqa
Comparing (\ref{target}) with the target space metric of ref.\cite{BCCGSW}, eq.(77), 
the translation rules are: ${\bf g}\rightarrow \lambda$, $e\rightarrow\tau^{\frac12}$,
$\varphi\rightarrow \mu$ and $\vec{A}\rightarrow \sqrt{3}\psi$ and $\vec{\psi}\rightarrow V$.

To reveal its group theoretical structure, we introduce a set of $G_{2(+2)}$ generators
$h_1$,$h_2$,$E^i_{~j}$ $(1\leq i\neq j\leq 3)$, $E^i$ $(1\leq i \leq 3)$ and
$E^{*}_i$ $(1\leq i \leq 3)$, satisfying \cite{Germar}
\begin{eqnarray}
{[}h_i ,h_j {]}&=&0, \nonumber\\
{[}h_i ,E^j_{~k} {]}&=&
\delta_i^j E^i_{\;\;k} - \delta_{i+1}^j E^{i+1}_{~~~~k}
-\delta_k^i E^j_{\;\;i} + \delta_k^{i+1} E^j_{\;\;i+1}, 
\nonumber\\
{[}h_i ,E^j {]}&=& \delta_i^j E^i
                     -\delta_{i+1}^j E^{i+1},\nonumber\\
{[}h_i ,E^*_j {]}&=& -(\delta_j^i E^*_i
                     -\delta_j^{i+1}E^*_{i+1}),\nonumber\\
{[}E^i_{\;\;j} ,E^k_{~l} {]}&=&
                     \delta^k_j E^i_{\;\;l}-\delta^i_l 
E^k_{\;\;j}, \nonumber\\
{[}E^i_{\;\;j} ,E^k {]}&=&
                     \delta^k_j E^i, \nonumber\\
{[}E^i_{\;\;j} ,E^*_k {]}&=&
                     -\delta^i_k E^*_j, \nonumber\\
{[}E^i ,E^j {]}&=&
                     -2\sum_k^3 \epsilon^{ijk}
                     E^*_k, \nonumber\\
{[}E^*_i ,E^*_j {]}&=&
                     +2\sum_k^3\epsilon_{ijk}
                     E^k, \nonumber\\
{[}E^i ,E^*_j {]}&=&
                     3 E^i_{\;\;j}
                     \hskip 2ex\mbox{if $i\neq j$},\nonumber\\
{[} E^1 ,E^*_1 {]}&=&2 h_1+ h_2,\nonumber\\
{[} E^2 ,E^*_2 {]}&=&- h_1+ h_2,\nonumber\\
{[} E^3 ,E^*_3 {]}&=&- h_1-2 h_2,
\label{G2generatorrelations}
\end{eqnarray}
where $\epsilon^{ijk}$ and $\epsilon_{ijk}$ are totally antisymmetric tensors with
$\epsilon^{123}=\epsilon_{123}=+1$ (The sign convention for $\epsilon_{123}$ was  
$-1$ in \cite{Germar}). 
This is the realization due to Fruedenthal, which shows the close relationship 
between the two exceptional Lie groups $E_8$ and $G_2$ \cite{MizoguchiE10,Germar}.
Then the $G_{2(+2)}$ group element
\beqa
{\cal V}^{(3)}&=&\exp\left(
-(\log e_1^{\bar 1}) h_1 - (\log e_1^{\bar 1}e_2^{\bar 2}) h_2 \right)
\exp(- e_1^{\bar 2}e_{\bar 2}^2 E^1_{~2})
\exp(\psi_i E^i_{~3})
\exp\left(-\frac1{\sqrt{3}}A_i E^i\right)
\exp\left(\frac1{\sqrt{3}}\varphi E^*_3\right)
\eeqa 
gives rise to a right invariant vector field 
\beqa
\partial_m {\cal V}^{(3)}{\cal V}^{(3)-1}&=&\frac12 \partial_m(\phi+\log\rho)h_1
+\frac12 \partial_m(\phi-\log\rho)h_2
-e^{\frac\phi 2}\rho^{\frac32}\partial_m B_z~E^1_{~2}
-\frac1{\sqrt{3}} e_{\bar a}^{~i}\partial_m A_i~E^{\bar a} \label{MC}
\\
&&+e^{-1}\left(
\frac1{\sqrt{3}}\partial_m\varphi-\frac13\epsilon^{ij}A_i\partial_m A_j
\right)E^*_3
+e^{-1}e_{\bar a}^{~i}\left(
\partial_m\psi_i-A_i\partial_m\varphi+\frac1{3\sqrt{3}}\epsilon^{kl}A_iA_k\partial_m A_l
\right)E^{\bar a}_{~3}.
\nonumber
\eeqa
To obtain the reduced Lagrangian (\ref{3DscalarLagrangian}), we define the 
symmetric space involution $\tau$, which is an automorphism and decomposes 
the $G_{2(+2)}$ Lie algebra into its eigenspaces:
\beqa
G_{2(+2)}&=&\bbsmall{H}\oplus \bbsmall{K},\n
\bbsmall{H}&=&\{X\in G_{2(+2)}~|~\tau(X)= +X \},\n
\bbsmall{K}&=&\{X\in G_{2(+2)}~|~\tau(X)= -X \}.
\eeqa
In the present case, $\bbsmall{H}$ is defined to be a subspace spanned by 
\beqa
E^1_{~2}+E^2_{~1},~E^1_{~3}+E^3_{~1},~
E^2_{~3}-E^3_{~2},~E^1-E^*_1,~E^2+E^*_2~\mbox{and}~E^3+E^*_3,
\eeqa
and $\bbsmall{K}$ is by
\beqa
E^1_{~2}-E^2_{~1},~E^1_{~3}-E^3_{~1},~
E^2_{~3}+E^3_{~2},~E^1+E^*_1,~E^2-E^*_2,~E^3-E^*_3,~h_1~\mbox{and}~h_2.
\eeqa
The reduced Lagrangian (\ref{3DscalarLagrangian})
can be obtained (with a suitable overall constant normalization factor) by 
projecting (\ref{MC}) onto $\bbsmall{K}$ and taking trace of the square. 
$\bbsmall{H}$ is the Lie algebra of the $SO(2,2) \sim SL(2,\RRsmall)\times SL(2,\RRsmall)$
subgroup, and hence $\bbsmall{K}$ the Lie algebra of the coset space $G_{2(+2)}/SO(2,2)$.
According to the general prescription, the coset representative is
\beqa
{\cal R}^{(3)}&=&\tau({\cal V}^{(3)-1}){\cal V}^{(3)},
\eeqa 
then $E^{(3)}\mbox{Tr}\partial_m{\cal R}^{(3)-1}\partial^m{\cal R}^{(3)}$ is automatically 
proportional to (\ref{3DscalarLagrangian}). Note that the coset representative can be chosen, as 
in \cite{BCCGSW}, to be a symmetric matrix, which can be obtained by multiplying a suitable constant matrix to ${\cal R}^{(3)}$ from, say,  the right, but it is not necessary.

The timelike-Killing $SL(2,\RRsmall)$ we considered in this paper acts as a
group multiplication to the $SL(2,\RRsmall)$ generated by $E^2$, $E^*_2$ and $-h_1+h_2$.
On the other hand, using the dictionalry given below (\ref{target}), one can identify that these 
generators are represented in \cite{BCCGSW} as
% They are actually E^1,E^*_1 and 2h_1+h_2 defined in G2_Gal'tsov.nb.
\beqa
E^2=\left(
\begin{array}
{ccccccc}
0&0&0&0&0&0&\sqrt{2}\\
0&0&0&0&0&0&0\\
0&0&0&0&0&0&0\\
0&0&0&0&0&0&0\\
0&0&1&0&0&0&0\\
0&-1&0&0&0&0&0\\
0&0&0&\sqrt{2}&0&0&0
\end{array}
\right),
E^*_2=\left(
\begin{array}
{ccccccc}
0&0&0&0&0&0&0\\
0&0&0&0&0&-1&0\\
0&0&0&0&1&0&0\\
0&0&0&0&0&0&\sqrt{2}\\
0&0&0&0&0&0&0\\
0&0&0&0&0&0&0\\
\sqrt{2}&0&0&0&0&0&0
\end{array}
\right),
h_2-h_1=
\left(
\begin{array}
{ccccccc}
2&0&0&0&0&0&0\\
0&-1&0&0&0&0&0\\
0&0&-1&0&0&0&0\\
0&0&0&-2&0&0&0\\
0&0&0&0&1&0&0\\
0&0&0&0&0&1&0\\
0&0&0&0&0&0&0
\end{array}
\right),
\eeqa
and the matrix ${\cal C}$ in \cite{BCCGSW} for the Harrison rotation 
\beqa
{\cal C}=
\left(
\begin{array}
{ccccccc}
c^2&0&0&s^2&0&0&\sqrt{2}sc\\
0&c&0&0&0&s&0\\
0&0&c&0&-s&0&0\\
s^2&0&0&c^2&0&0&\sqrt{2}sc\\
0&0&-s&0&c&0&0\\
0&s&0&0&0&c&0\\
\sqrt{2}sc&0&0&\sqrt{2}sc&0&0&c^2+s^2
\end{array}
\right),~~~c=\cosh\alpha,~s=\sinh\alpha
\eeqa
can be written as $\exp(-\alpha(E^2+E^*_2))$, and hence belong to the 
timelike-Killing $SL(2,\RRsmall)$. This explains why we have obtained the 
Cveti{\v c}-Youm solution in Appendix A. Note that  $\exp(-\alpha(E^2+E^*_2))$ 
is not the same as the $SL(2,\RRsmall)$ group element used in Appendix A;
this redundancy of the $SL(2,\RRsmall)$ group element was already reported 
in the spacelike Killing case \cite{MT}.


\begin{thebibliography}{99}

\bibitem{ER}
R.~Emparan and H.~S.~Reall,
Phys.\ Rev.\ Lett.\  {\bf 88}, 101101 (2002).
\bibitem{Pom}
A. A. Pomeransky and R.A. Sen'kov, e-Print: arXiv:hep-th/0612005.
\bibitem{MishimaIguchi}
T.~Mishima and H.~Iguchi, Phys. Rev. D {\bf 73}, 044030 (2006).
\bibitem{EEMR}
H. Elvang, R. Emparan, D. Mateos and H. S. Reall,
Phys. Rev. Lett. {\bf 93}, 211302 (2004).
\bibitem{EEMR2}
H. Elvang, R. Emparan, D. Mateos and H. S. Reall, Phys. Rev. D {\bf 71}, 024033 (2005).
\bibitem{Y3}
S. S. Yazadjiev, Phys. Rev. D {\bf 73}, 104007 (2006).



\bibitem{diring} 
H. Iguchi and T. Mishima, Phys. Rev. D {\bf 75}, 064018 (2007).
\bibitem{saturn}
H. Elvang and P. Figueras, JHEP {\bf 0705}, 050 (2007).
\bibitem{Izumi}
K. Izumi, Prog. Theor. Phys. {\bf 119}, 757 (2008).
\bibitem{bi}
H. Elvang and M. J. Rodriguez, JHEP {\bf 0804}, 045 (2008).


\bibitem{EEF}
H. Elvang, R. Emparan and P. Figueras, JHEP {\bf 0502}, 031 (2005).





\bibitem{MO}
S.~Mizoguchi and N.~Ohta,
  Phys.\ Lett.\  B {\bf 441}, 123 (1998).
\bibitem{BCCGSW}
A. Bouchareb, G. Clement, C-M. Chen, D. V. Gal'tsov, N. G. Scherbluk and T. Wolf, Phys. Rev. D {\bf 76},104032 (2007); Erratum-ibid. D {\bf 78}, 029901 (2008).
\bibitem{TYI}
S. Tomizawa, Y. Yasui and A. Ishibashi, Phys. Rev. D {\bf 79}, 124023 (2009).
\bibitem{TYI2}
S. Tomizawa, Y. Yasui and A. Ishibashi, Phys. Rev. D {\bf 81}, 084037 (2010).
\bibitem{T}
S. Tomizawa, Phys. Rev. D {\bf 82}, 104047 (2010).
\bibitem{GRS}
S. Giusto,  S. F. Ross and  A. Saxena, JHEP {\bf 12}, 065 (2007). 






\bibitem{CBJV}
G. Compere, S. Buyl, E. Jamsin and A. Virmani, Class. Quant. Grav. {\bf 26}, 125016 (2009).
\bibitem{CBSV}
G. Compere, S. Buyl, S. Stotyn and A. Virmani, JHEP {\bf 11}, 13 (2011).
\bibitem{TYM}
S. Tomizawa, Y. Yasui and Y. Morisawa, Class. Quant. Grav. {\bf 26}, 145006 (2009).
\bibitem{GS1}
D. V. Gal'tsov and N. G. Scherbluk, Phys. Rev. D {\bf 78}, 064033 (2008).
\bibitem{GS2}
D. V. Gal'tsov and N. G. Scherbluk, Phys. Rev. D {\bf 79}, 064020 (2009).


\bibitem{CN}
  A.~H.~Chamseddine and H.~Nicolai,
  Phys.\ Lett.\  B {\bf 96}, 89 (1980).
\bibitem{Germar}  
 S.~Mizoguchi and G.~Schr\"oder,
  Class.\ Quant.\ Grav.\  {\bf 17}, 835 (2000).
  
\bibitem{MizoguchiAsymObfd}
  S.~Mizoguchi,
  %``On asymmetric orbifolds and the D = 5 no-modulus supergravity,''
  Phys.\ Lett.\  B {\bf 523}, 351 (2001)
  [arXiv:hep-th/0109193].




\bibitem{MT}
S. Mizoguchi and S. Tomizawa, Phys. Rev. D {\bf 84}, 104009 (2011).



\bibitem{Tomizawa}
S. Tomizawa, Phys. Rev. D {\bf 82}, 104047 (2010).






\bibitem{DM}
P. Dobiasch and D. Maison, Gen. Rel. Grav. {\bf 14} (1982), 231. 
\bibitem{Rasheed}
D. Rasheed, Nucl. Phys. B {\bf 454}, 379 (1995). 
\bibitem{IM}
H. Ishihara and K. Matsuno, Prog. Theor. Phys. {\bf 116}, 417 (2006).
\bibitem{NIMT}
T. Nakagawa, H. Ishihara, K. Matsuno and S. Tomizawa, Phys. Rev. D {\bf 77}, 044040 (2008). 
\bibitem{TIMN}
S. Tomizawa, H. Ishihara, K. Matsuno and T. Nakagawa, Prog. Theor. Phys. {\bf 121}, 823 (2009).
\bibitem{TI}
S. Tomizawa and A. Ishibashi, Class. Quant. Grav. {\bf 25}, 245007 (2008).
\bibitem{T2}
S. Tomizawa, e-Print: arXiv:1009.3568 [hep-th].
\bibitem{Gauntlett0}
J. P. Gauntlett, J. B. Gutowski, C. M. Hull, S. Pakis and H. S. Reall, Class. Quant. Grav. {\bf 20}, 4587 (2003). 
\bibitem{Gaiotto}
D. Gaiotto, A. Strominger and X. Yin, JHEP {\bf 02}, 023 (2006).
\bibitem{Elvang3}
H. Elvang, R. Emparan, D. Mateos and H. S. Reall, JHEP {\bf 08} (2005), 042













\bibitem{Ehlers}
 J. Ehlers, Dissertation, Univ. Hamburg (1957).
 
 \bibitem{MM}
 R.~A.~Matzner and C.~W.~Misner,
  Phys.\ Rev.\  {\bf 154}, 1229 (1967).


\bibitem{BMG}
  P.~Breitenlohner, D.~Maison and G.~W.~Gibbons,
  %``Four-Dimensional Black Holes from Kaluza-Klein Theories,''
  Commun.\ Math.\ Phys.\  {\bf 120}, 295 (1988).
  %%CITATION = CMPHA,120,295;%%


\bibitem{MP}
R.~C.~Myers and M.~J.~Perry,
Annals Phys.\  {\bf 172}, 304 (1986).


\bibitem{CY96}
M. Cveti\v{c} and D. Youm, Nucl. Phys. B {\bf 476}, 118 (1996).


 \bibitem{CJ}
 E.~Cremmer and B.~Julia,
  Nucl.\ Phys.\  B {\bf 159}, 141 (1979).


\bibitem{weyl}  
R. Emparan and H. S. Reall, Phys. Rev. D {\bf 65}, 084025 (2002).
%\bibitem{TYI}
%S. Tomizawa, Y. Yasui and A. Ishibashi, Phys. Rev. D{\bf 79}, 124023 (2009).





  
 \bibitem{MizoguchiE10}
  S.~Mizoguchi,
  Nucl.\ Phys.\  B {\bf 528}, 238 (1998).
  

  
  
  
  
  
  
  
  
  
  
  
  
  
\if0
  

\bibitem{Sch}
K. Schwarzschild, {\it Sitzber. Deut. Akad. Wiss. Berlin}, KL. Math.-Phys. Tech. 189 (1916).

\bibitem{p-brane} 
  M.~J.~Duff, R.~R.~Khuri and J.~X.~Lu, Phys.\ Rept.\  {\bf 259}, 213 (1995).

\bibitem{SV}
 A.~Strominger and C.~Vafa,
  Phys.\ Lett.\  B {\bf 379}, 99 (1996).





\bibitem{Elvang3}
H. Elvang, R. Emparan, D. Mateos and H. S. Reall, JHEP {\bf 08}, 042 (2005).
\bibitem{BKW}
I. Bena, P. Kraus and N.P. Warner, Phys. Rev. D {\bf 72}, 084019 (2005). 
\bibitem{Bena}
I. Bena and P. Kraus, Phys. Rev. D {\bf 70}, 046003 (2004). 
\bibitem{Bena2}
I. Bena, P. Kraus and R. Warner, Phys. Rev. D {\bf 72}, 084019 (2005).
\bibitem{Bena3}
I. Bena and N. P. Warner, Adv. Theor. Math. Phys. {\bf 9}, 667 (2005).
\bibitem{BGRW}
I. Bena, S. Giusto, C. Ruef and N. P. Warner, JHEP {\bf 11}, 032 (2009).

\bibitem{FGPS}
J. Ford, S. Giusto, A. Peet and A. Saxena, Class. Quant. Grav. {\bf 25}, 075014 (2008).
\bibitem{CEFGS}
J. Camps, R. Emparan, P. Figueras, S. Giustod and A. Saxena, JHEP {\bf 02}, 021 (2009).

\bibitem{Gibbons-Perry}
G. W. Gibbons and M. J. Perry, Nucl. Phys. B {\bf 248}, 629 (1984).
\bibitem{CY-KK3}
M. Cveti{\v c} and D. Youm, Nucl. Phys. B {\bf 438}, 182 (1995).
\bibitem{CY-KK4}
M. Cveti{\v c} and D. Youm, Nucl. Phys. B {\bf 453}, 259 (1995).
\bibitem{CY-KK5}
M. Cveti{\v c} and D. Youm, Phys. Rev. D {\bf 52}, 2574 (1995).
\bibitem{Nelson}
W. Nelson, Phys. Rev. D {\bf 49}, 5302 (1994). 


\bibitem{Herdeiro0}
C. A. R. Herdeiro, Nucl. Phys. B {\bf 665}, 189 (2003).
\bibitem{Gimon-Hashimoto}
E. Gimon and A. Hashimoto, Phys. Rev. Lett. {\bf 91}, 021601 (2003).
\bibitem{Herdeiro}
C. A. R. Herdeiro, Class. Quant. Grav. {\bf 20}, 4891 (2003).
\bibitem{Wu}
S-Q. Wu, Phys. Rev. Lett. {\bf 100}, 121301 (2008).
\bibitem{MINT}
K. Matsuno, H. Ishihara, T. Nakagawa and S. Tomizawa, Phys. Rev. D {\bf 78}, 064016 (2008). 

  
 \bibitem{Geroch}
 R.~P.~Geroch,
  J.\ Math.\ Phys.\  {\bf 12}, 918 (1971); 
  J.\ Math.\ Phys.\  {\bf 13}, 394 (1972).
  
 \bibitem{CFS}
 E.~Cremmer, J.~Scherk and S.~Ferrara,
  Phys.\ Lett.\  B {\bf 74}, 61 (1978).
  
 

  
  \bibitem{Cremmer}
  E. Cremmer, in: ?gCambridge 1980, Proceedings, Superspace and Supergravity?h, eds.
S. W. Hawking and M. Roc{\v e}k (Cambridge University Press, 1981) 267.

  
  

  

 \bibitem{CJS}
 E.~Cremmer, B.~Julia and J.~Scherk,
  Phys.\ Lett.\  B {\bf 76}, 409 (1978).
 
 \bibitem{HT}
 C.~M.~Hull and P.~K.~Townsend,
  Nucl.\ Phys.\  B {\bf 438}, 109 (1995).
  
 \bibitem{OP}
 N.~A.~Obers and B.~Pioline,
  Phys.\ Rept.\  {\bf 318}, 113 (1999).
 
 \bibitem{MarcusSchwarz}
 N.~Marcus and J.~H.~Schwarz,
  Nucl.\ Phys.\  B {\bf 228}, 145 (1983).
  











\bibitem{Giusto-Saxena}
S. Giusto and A. Saxena, Class. Quantum Grav. {\bf 24}, 4269 (2007).


\bibitem{BMPV}
J. C. Breckenridge, R.C. Myers, A.W. Peet and C. Vafa, Phys. Lett. B {\bf 391}, 93 (1997).



\bibitem{Rasheed-Gibbons1}
G. W. Gibbons and D. A. Rasheed, Nucl. Phys. B {\bf 454}, 185 (1995).
\bibitem{Rasheed-Gibbons2}
G. W. Gibbons and D. A. Rasheed, Phys. Lett. B {\bf 365}, 46 (1996).


%%%%%%%%%%%%%%%%%%%%




\bibitem{NicolaiSchladming}
gTwo-dimensional Gravities and Supergravities as Integrable Systemh,
30th Schladming Winter School, Schladming (1991).


\fi
 
\end{thebibliography}
\end{document}